\documentclass[journal]{IEEEtran}
\usepackage{cite}
\usepackage{multirow}
\usepackage{graphicx,amsmath,algorithm,algorithmic,array}
\usepackage{color}

\begin{document}
\title{Latency Guarantee for Ubiquitous Intelligence in 6G: A Network Calculus Approach}
\author{Lianming Zhang, Qian Wang, Pingping Dong, Yehua Wei, and Jing Mei
\thanks{Lianming Zhang, Qian Wang, Pingping Dong, Yehua Wei, and Jing Mei are with the Hunan Normal University of China.
}
\thanks{Manuscript received May 5, 2022.}}


\maketitle

\begin{abstract}
With the gradual deployment of 5G and the continuous popularization of edge intelligence (EI), the explosive growth of data on the edge of the network has promoted the rapid development of 6G and ubiquitous intelligence (UbiI). This article aims to explore a new method for modeling latency guarantees for UbiI in 6G given 6G's extremely stochastic nature in terahertz (THz) environments, THz channel tail behavior, and delay distribution tail characteristics generated by the UBiI random component, and to find the optimal solution that minimizes the end-to-end (E2E) delay of UbiI. In this article, the arrival curve and service curve of network calculus can well characterize the stochastic nature of wireless channels, the tail behavior of wireless systems and the E2E service curve of network calculus can model the tail characteristic of the delay distribution in UbiI. Specifically, we first propose demands and challenges facing 6G, edge computing (EC), edge deep learning (DL), and UbiI. Then, we propose the hierarchical architecture, the network model, and the service delay model of the UbiI system based on network calculus. In addition, two case studies demonstrate the usefulness and effectiveness of the network calculus approach in analyzing and modeling the latency guarantee for UbiI in 6G. Finally, future open research issues regarding the latency guarantee for UbiI in 6G are outlined.
\end{abstract}

\IEEEpeerreviewmaketitle

\section{Introduction}
With 5G entering the stage of commercial deployment, research institutions around the world have begun to focus on 6G. The goal of 6G is to meet the needs of the intelligent information society in the 2030s and form an autonomous ecosystem with human wisdom and consciousness \cite{Strinati2019}. Artificial intelligence (AI) research aims to extend and enhance human ability and efficiency in various tasks of transforming nature and governing society through intelligent machines to ultimately achieve a society in which humans and machines coexist harmoniously. The design of the 6G network architecture should follow the AI-driven method, in which intelligence will become an inherent characteristic of the 6G network architecture. It is foreseeable that the deep integration of 6G and AI will form a wireless intelligent (WI) environment with a ubiquitous network and intelligence \cite{Latva-aho2019}.

WI and the Internet of Things (IoT) collaborate to promote the rapid development of AI of Things (AIoT). The explosive growth of data on the edge side of the AIoT has promoted the development of edge computing (EC) \cite{Zhang2018}. EC provides intelligent services on the edge side of the AIoT near the source of the data and meets the key requirements of industry digitalization in terms of real-time business and application intelligence. With the rapid development of AI, especially deep learning (DL), edge intelligence (EI) has emerged. EI aims to efficiently deploy DL models in resource-constrained terminal devices, making intelligence closer to users \cite{Zhou2019}.

\begin{figure}[htbp]
\centerline{\centering
\includegraphics[width=3.15in]{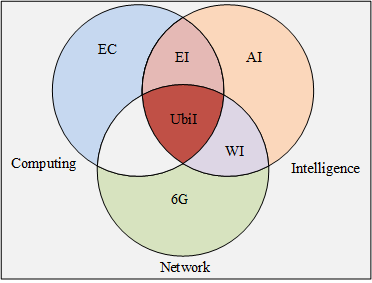}
}
\caption{UbiI, 6G, EC and AI. }
\label{fig1}
\end{figure}

The natural fusion of WI and EI produces ubiquitous intelligence (UbiI) in which the "network is everywhere, computing power is everywhere, and intelligence is everywhere", as shown in Fig. \ref{fig1}.

In UbiI systems, delay-sensitive services (such as the industrial Internet and fully automatic driving) require a strict and extremely low delay. 6G is expected to use terahertz (THz) technology, which will greatly increase the network capacity and network speed of 6G networks and provide extremely low latency \cite{Zhang2019}. However, the THz environment has high variability and uncertainty \cite{Saad2020}, and random traffic demand, delay, interference, and deep fading in wireless systems all have tail behavior \cite{Bennis2018}. These severely affect the worst-case delay. EC shortens the distance between the application and the server and reduces the over-the-air communication delay, which contributes greatly to the end-to-end (E2E) delay. However, the densification of the network and the mobility of users brought about by EC will increase the E2E delay. Edge DL can generate inferences in edge device instances, thereby providing an opportunity to greatly reduce E2E latency in EC applications. However, edge DL devices cannot access large global training datasets. To improve inference accuracy, edge DL devices need to cooperate, which will generate additional overhead and increase training delay. In particular, in the UbiI system where 6G, EC, and edge DL are integrated, queuing delay, computing delay and communication delay are all stochastic. The high variability and uncertainty of the 6G environment, the tail behavior of the wireless system, and the randomness of delays in EC and DL will inevitably lead to tail behavior of the delay distribution in the UbiI system.

Existing average-based network design approaches rely on average quantities such as throughput, latency, and response time and cannot characterize the tail characteristics of the delay distribution \cite{Bennis2018}. Therefore, a principled framework that characterizes the stochastic nature and tail behavior in the THz environment is needed to model latency guarantees for UbiI in 6G. In particular, it should be able to measure the worst-case E2E delay and other performance indicators. This article uses a network calculus approach to model latency guarantees for UbiI in 6G. In particular, the arrival curve and service curve of network calculus can well characterize the high variability and uncertainty of the THz environment and tail behavior of 6G wireless systems, and the E2E service curve of network calculus characterizes the worst-case E2E delay in EC, and DL will inevitably lead to the tail behavior of the delay distribution in the UbiI system.

In this article, we first discuss the demands and challenges of 6G, EC, edge DL, and UbiI in latency guarantees. Then, we propose feasible solutions, including the hierarchical architecture of the UbiI system, the UbiI network model, and latency guarantee for the UbiI. In addition, two case studies validate the effectiveness of the delay model based on network calculus. Finally, some open issues on the latency guarantee for UbiI in 6G are discussed for future work.

\section{Demands and Challenges}
In UbiI in 6G, the components that generate delay can be divided into deterministic components and statistical components \cite{Bennis2018}. Deterministic latency components consist of the time to check bit errors and determine the output link and the time to send bits into the link. The statistical latency components include the time waiting at the output link for transmission, propagation delays, and other processing and computing delays. To achieve ultralow latency, several technologies need to be implemented in 6G UbiI systems \cite{Elbamby2019}. First, the THz technology expected to be used in 6G provides a wide bandwidth with high directivity, which helps reduce the delay between the user equipment and the base station (BS). Second, EC is based on the idea of proximity, which shortens the data transmission distance and reduces the delay required for task offloading from the client device to the cloud center. Third, edge DL is essential for low-latency mission-critical applications by authorizing edge servers or edge devices to execute decisions locally. Finally, UbiI in 6G will integrate the abovementioned technologies to achieve a principled and scalable framework for ultralow latency and decision making under uncertainty.

\subsection{6G}
6G's key performance indicators, such as peak transmission rate, communication delay, and ultrahigh density, will be 10-100 times higher than 5G \cite{Latva-aho2019}. However, the densification of 6G networks places strict requirements on area or space spectrum efficiency, as well as the frequency band required for connection. THz communication with a frequency band of 100 GHz to 10 THz is considered to be the most important technology for 6G mobile communication. The higher the frequency is, the shorter the wavelength, the worse the diffraction ability of the signal, the shorter the transmission distance, and the greater the loss. This loss will increase with the transmission distance, and the coverage of the BS will decrease accordingly. THz-level 6G signals are easily absorbed by water molecules in the air, and the signal loss is very serious. The coverage area of the BS will be smaller, and the density of the BS will increase. Therefore, THz-level 6G will face severe challenges of improving coverage, reducing interference, and solving deep fading. Additionally, solving the inherent tail behavior of wireless systems is a long-standing challenge. This tail behavior is inherently related to the tail of random traffic demand, the tail behavior of delay distribution, interference, limited functionality, and deep fading.

\subsection{EC}
EC shortens the distance between the client device and the edge server and reduces the communication delay \cite{Elbamby2019}. 6G provides opportunities for the introduction of other computing resources at the edge of the network while increasing capacity and coverage. The user-end device uploads its computing task to the edge server and downloads the corresponding output after high data rate processing, which reduces the computing delay and communication delay. Due to the limited availability of EC communications and computing resources, the increased interference of network density may reduce the quality of uplink and downlink communications, thereby increasing E2E delay. Furthermore, the uncertainty of the EC environment and the mobility of user equipment may cause undesirable delays. Consequently, using a high-frequency THz 6G channel to reduce computing delay and communication delay is the main challenge of EC as a low-latency enabler between task offloading from client devices to the cloud center.

\subsection{Edge DL}
Training data of edge DL are stored in a distributed manner on multiple interconnected edge devices, which can generate inferences in the instances of edge devices, thereby providing an opportunity to greatly reduce the E2E latency in mobile EC applications \cite{Elbamby2019,Samarakoon2018}. Edge DL can predict the channel dynamics, communication, and computing resource availability of local devices. Even if the connection is interrupted, the edge DL deployed on the client device allows predicting system behavior and making decisions within the device, reducing the number of parallel tasks performed by the edge server and sequentially reducing task input and output delays. Since edge devices cannot access large global training datasets, it will reduce the inference accuracy of local data training. For this reason, edge devices may often need to collaborate with each other or a centralized assistant. However, this will generate additional overhead and increase the training delay. Therefore, optimizing the tradeoff between training delay and inference accuracy is the main challenge for edge DL as a low-latency enabler for mission-critical applications.

\subsection{UbiI}
UbiI is the integration of multiple technologies of 6G, EC, and edge DL. The high-capacity THz link in 6G greatly reduces the communication delay between the user terminal and the BS. However, the tail behavior of 6G wireless systems and the uncertainty of the THz environment have long existed. Currently, commonly used network design methods based on average quantities cannot solve these fine-grained indicators, such as the behavior of the delay distribution tail. Edge computing uses proximity-based ideas to reduce the E2E air communication delay that makes a significant contribution by shortening the distance between the client device and the edge server. Nonetheless, intermittent congestion and interruption of the THz channel and user mobility always exist in EC, and many factors affect latency at the network level. Edge DL provides low-latency inference (or prediction) functions, but edge devices cannot access large global training data sets, and the uncertainty of interference and network congestion always exists. In particular, in EC and edge DL, communication delays, computing delays, and queuing delays required for communication and computing are still encountered. UbiI systems need to shift from an average-based network design to a clean-slate design centered on the tail. Understanding the tail behavior of UbiI systems and designing a principled and extensible framework for describing these tail characteristics is a long-term challenge.

\section{Solutions}
In this section, we propose a solution to guarantee the low latency of the UbiI system based on network calculus. First, we propose the hierarchical architecture of the UbiI system. Then, we build a network model based on UbiI. Finally, we use network calculus to derive UbiI's E2E delay.

\subsection{Hierarchical architecture of UbiI system}
UbiI is a kind of network-based intelligence of all things, and it is the perfect combination of the network, computing power, and intelligence. There are two meanings: EI supported by the 6G network and the 6G network driven by EI.

Many manufacturers are now integrating DL functions into IoT devices, making IoT devices increasingly intelligent. UbiI is expected to fully exploit the potential of large-scale IoT devices and data generated at the edge of the network to support intelligent applications with lower latency. Regarding the EI stratification method in \cite{Zhou2019}, we divide the UbiI architecture into four levels from low to high: (1) cloud training and inference, edge/device inference, (2) cloud training, edge/device inference, (3) cloud training and inference, edge/device training and inference, and (4) edge/device training and inference.

\subsection{UbiI network model}
\begin{figure*}[htbp]
\centerline{\centering
\includegraphics[width=5in]{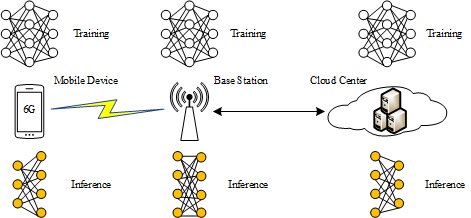}
}
\caption{UbiI network model.}
\label{fig2}
\end{figure*}

Edge DL deployed in the UbiI system maps each part of the deep neural networks (DNN) and other neural networks \cite{Zhang2020} to different computing devices to minimize the device's communication and resource usage, and it is also an important method to reduce delay. According to the application requirements of the UbiI system, an appropriate UbiI level is selected and cooperates to complete training and inference in the cloud, edge, and devices. After the DNN is trained, the cloud center, edge server, and mobile device will be distributed and executed according to the obtained scheme. For example, if the third level is selected, the cloud center, edge server, and mobile device use the DNN to extract and learn these training data and their optimal scheme features. Fig. \ref{fig2} shows the UbiI system network model including a cloud center, a BS, and a mobile device. A wired (such as optical fiber) connection is adopted between the cloud center and the BS, and 6G communication is directly used between the BS and the mobile device.

In a UbiI system, for a given predetermined number of cloud centers, BSs, and mobile devices, it is necessary to find the path of task/data unloading with the minimum service delay and the location deployment of the associated cloud center, BSs, and mobile devices. Similar to Ref. \cite{Koketsu2019}, it is assumed that each mobile device is connected to only one BS, and one BS can connect zero or more mobile devices. Each BS is only connected to one cloud center, and one cloud center can be connected to zero or more BSs. Assuming that one or more mobile devices periodically generate tasks and data to be processed, the mobile device can either queue for processing tasks and data with one processor or send these tasks and data to neighboring BSs through a 6G wireless link. Then, the BS can use one of the processors to queue the entire task and data, send a part of the task and data to the cloud center through a wired backhaul link, or send a part of the task and data to mobile devices adjacent to the BS through a 6G wireless link. Once the cloud center and mobile device receive these tasks/data, one of the processors is also used to queue tasks and data. When one processor in the BS, cloud center, and mobile device is in the idle state, the DNN is used to extract and learn these training data and their optimal scheme characteristics. After the DNN is trained, the BS, cloud center, and mobile device perform distributed execution and generate results according to the obtained scheme. Then, the results are sent back to the mobile device that originally sent the task through the direct path taken by the task/data.

\subsection{Latency guarantee for UbiI}
Service delay is the interval between the task/data generation and the result return of the mobile device. It can be divided into two stages: communication delay and computing delay. Communication delay mainly includes the time on the wireless channel of the 6G communication between the mobile device and the BS and the transmission time of the backhaul link between the BS and cloud center. The total communication delay consists of the time required to transmit tasks in the uplink and the time required to transmit results in the downlink. This delay includes two transmission delays between the mobile device and the associated BS and between the associated BS and the associated cloud center. The communication delay between the mobile device and the associated BS is modeled by wired network calculus \cite{Boudec2004}, and the communication delay between the associated BS and the management cloud center is modeled by wireless network calculus \cite{Al-Zubaidy2016}. The computing delay mainly includes the time that the task waits for the processor to be idle in the cloud, the BS, the mobile device and the time for DNN model training and inference. Training delay is captured by loss or weight convergence, while inference delay refers to the computing and memory access delay \cite{Park2019}. The computing and training delay is also modeled by network calculus.

\begin{figure*}[htbp]
\centerline{\centering
\includegraphics[width=5in]{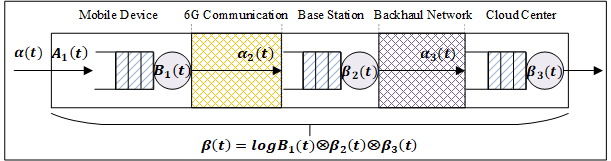}
}
\caption{Service delay model.}
\label{fig3}
\end{figure*}

To obtain the latency guarantee for the UbiI system, we must minimize the E2E service delay modeled by network calculus, as shown in Fig. \ref{fig3}. Here, $A_{1} (t)$ and $B_{1} (t)$ represent the arrival curve and service curve, respectively, of the mobile device in the signal-to-noise ratio (SNR) domain, $\alpha _2 (t)$, $\beta _2 (t)$, $\alpha _3 (t)$, and $\beta _3 (t)$ represent the arrival curve and service curve of the BS and cloud center in the bit domain, and $\alpha (t)$ and $\beta (t)$ represent the arrival curve and service curve of the whole system in the bit domain, respectively. According to wireless network calculus, the arrival curve and the service curve in the SNR domain are exponential functions of the arrival curve and service curve in bit domain, respectively, and the total service curve $\beta (t)$ of a series system can be regarded as the convolution $\otimes$ of single system service curves. Therefore, the E2E delay of the UbiI system can be obtained, and the service time based on the backhaul time and the number of communication rounds is derived.

\section{Case studies}
To illustrate the usefulness and effectiveness of the network calculus approach in analyzing and modeling the latency guarantee for UbiI in 6G, we focus our attention on two use cases on different traffic demands. We assume that the data generated on mobile devices are constrained by the curve. The training and inference provided by mobile devices, BSs, and cloud centers are described using service curves. We consider the two most common combinations of arrival curves and service curves.

\subsection{Case 1}
To control the rate of data injected into the UbiI system and smooth the burst traffic on the UbiI, leaky buckets are used to shape the traffic or limit the rate. Through leaky buckets, burst traffic can be shaped to provide stable traffic for the UbiI. The traffic constrained by the arrival curve of the leaky bucket is called leaky bucket traffic. Assuming a mobile device, BS, or cloud center provides a rate latency service curve. The leaky bucket traffic passes through a mobile device, BS, or cloud center, and the delay bound is obtained.

Since the mobile device and the BS communicate through 6G networks, it is necessary to map the leaky bucket traffic in the bit domain of the mobile device to the SNR domain. The arrival curve and service curve in the SNR domain are exponential functions of the arrival curve and service curve in the bit domain, respectively. Therefore, the arrival curve and service curve of the corresponding mobile device in the SNR domain can be determined. Using the related properties of $(min,\times)$ dioid algebra, the delay bound of the mobile device in the SNR domain can be derived.

The conversion between the bit domain and the SNR domain does not affect the delay. According to the arrival curve entering the mobile device, the service curves provided by the mobile device, BS, and cloud center, the E2E delay bound of the UbiI system can be calculated by using the convolution (E2E service curve) of the three rate latency service curves of the mobile device, BS, and cloud center.

\subsection{Case 2}
To increase perceived overall quality, the variable bit rate (VBR) is often used in encoding lossy formats. On the one hand, VBR traffic is constrained by the leaky bucket. On the other hand, VBR traffic is also limited by the network link capacity and the maximum length of data packets, so its arrival curve can be obtained. The VBR traffic passing through the rate latency service curve can be regarded as the traffic restricted by two tandem leaky buckets passing through the rate latency service curve of the mobile device, BS, and cloud center.

Similarly, the E2E delay bound of the UbiI system can be calculated using the convolution (E2E service curve) of the three rate delay service curves of the mobile device, BS, and cloud center \cite{Boudec2004}.

\begin{figure*}[htbp]
\centerline{\centering
\includegraphics[width=6in]{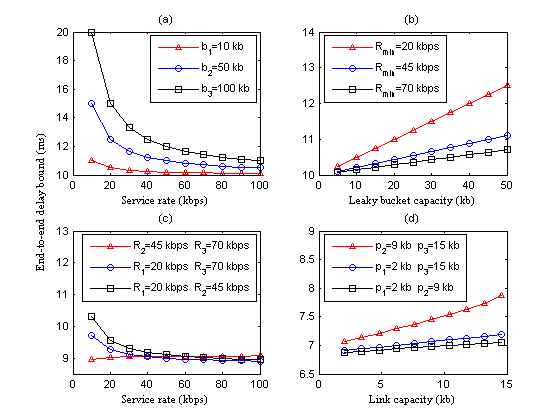}
}
\caption{E2E delay bound. (a) Delay bound versus service rate when fixing the leaky bucket capacity $b_{1}$, $b_{2}$, $b_{3}$; (b) delay bound versus leaky bucket capacity when fixing minimum service rate $R_{min}$; (c) delay bound versus service rate when fixing two of three service rates $R_{1}$, $R_{2}$, $R_{3}$; (d) delay bound versus link capacity when fixing two of three link capacities $p_{1}$, $p_{2}$, $p_{3}$.}
\label{fig4}
\end{figure*}

Figs. \ref{fig4}(a)-(b) and Figs. \ref{fig4}(c)-(d) show the relationship between the upper bound of E2E delay and computing power (service rate), computing capacity (leaky bucket capacity, and link capacity) in cases 1 and 2. As observed in Fig. \ref{fig4}(a), the upper bound of the E2E delay decreases exponentially with the decrease in the service rate, with larger changes obtained for greater leaky bucket capacity. As seen in Fig. \ref{fig4}(b), the upper bound of the E2E delay increases linearly with the increase in the leaky bucket capacity. A greater change is obtained for the smaller minimum service rate in mobile devices, BSs, and cloud centers. As seen in Fig. \ref{fig4}(c), when the service rate of the mobile device and the BS is constant, the upper bound of the E2E delay decreases exponentially with an increasing service rate of the cloud center. When the service rates of the mobile devices and cloud centers are constant, the E2E delay upper bound decreases exponentially with increasing service rates of the BSs. When the service rates of the BS and the cloud center are constant, the upper bound of the E2E delay increases slowly and reaches the equilibrium point with increasing service rates of mobile devices. It is seen in Fig. \ref{fig4}(d) that when the capacity of one of the 6G communication and backhaul network links is constant, the E2E delay upper bound increases linearly or approximately linearly with the increase in the capacity of the other link.

\subsection{Numerical results}

It can be seen from the above that the upper bound of the E2E delay of the UbiI system decreases exponentially with the increase in the service rates of the mobile devices, BSs, and cloud centers; it increases linearly or approximately linearly with the increase in the leaky bucket capacity and link capacity. The UbiI system service delay also follows this rule. Moreover, according to network calculus \cite{Al-Zubaidy2016}, we can obtain the arrival curve of the mobile device and the service curve in the SNR domain. Replacing the arrival curve and service curve of 6G communication in the UbiI system's E2E delay with the function of SNR, it is easy to deduce the influence of SNR on the service delay of the UbiI system in 6G communication.

\section{Research issues}
In this section, some open research issues of the latency guarantee for UbiI in 6G are discussed.

\subsection{6G ultralow latency}
From 2G to 5G, the delay of the mobile communication network centered on service personnel depends on the delays due to the human auditory ($\sim$100 ms), visual (approximately 10 ms), olfactory, tasteful, and tactile (1 ms) reaction time \cite{Yang2019}. Its design focuses on improving network capacity but pays little attention to delay \cite{Bennis2018}. 6G networks will provide stable ($<$1 ms) or even nonexistent delays, no longer just relying on the human response time. The open problems faced by the implementation of 6G ultralow latency include using terahertz or visible light to further compress the basic time-slot unit in 6G, improving the intermittent blocking and interruption of the THz channel, quickly capturing channel state information (CSI), and sensing directional mobility.

\subsection{EI delay}
EI systems are constrained by resources on communications and devices \cite{Park2019}. The size of the DNN and its energy consumption may exceed the device's memory size and battery capacity, hindering distributed inference. The process of decentralized training involves a large number of devices interconnected by wireless links. Due to the outdated model state information exchange under the condition of bad wireless channels, the training convergence is hindered. Using EI can improve communication, but it also increases the extra delay for inferring. Using a well-trained DL model and a large number of data samples, the training delay of the DL model is ignored. Therefore, under the condition of meeting the E2E delay and equipment hardware requirements, using DL at the network edge will give rise to new basic research problems for joint optimization of training, communication, and control.

\subsection{UbiI delay}
UbiI is affected not only by the delay caused by the hardware itself but also by the delay required for wired and wireless communications and computing. The low latency of wireless communication in UbiI depends on the following two aspects \cite{Bennis2018}. On the one hand, when the channel changes over time and there is uncertainty in the dynamics of the network, there will be risks when dealing with decisions and requirements to provide performance guarantees and robust decentralized or semicentralized DL algorithms. On the other hand, the tail behavior in the wireless system is inherently related to the tails of random traffic demand and delay distribution, so finding ways to accurately characterize the tails of these percentiles and extreme events remains an open issue.

\subsection{Mission-critical applications}
The main focus of the recent increase in mission-critical applications, such as industrial Internet and fully automatic driving, is to provide services with guaranteed high reliability and low latency.

The intelligent factory in the industrial Internet is composed of dense intelligent mobile robots that need wireless access to high-performance computing resources. The robot will need to respond quickly to changing conditions, including interaction with people, and operate in a time-critical control cycle. This massive wireless capacity will require an ultralow delay of less than 10 $\mu$s. Therefore, while 6G provides a large bandwidth for the industrial Internet to achieve the required data density, determining how to eliminate the stochastic nature of the wireless channel and the tail behavior in the wireless system on the extremely low latency of the industrial Internet is an open research issue.

Fully automatic driving means that all driving operations are completed by the unmanned driving system. When possible, humans take over without limiting road and environmental conditions. Automatic driving must quickly merge multisource data to decide how to control the vehicle. However, the progress of data-driven DL alone cannot solve the safety of fully automated driving approaching 100 percent, which requires different ideas. The DL system for E2E training in EI may be too complicated to allow engineers to separately test the quality of its components, which poses a problem for fully automated driving systems that require extremely high safety. Moreover, the mobility and wireless channel fading of fully automatic driving cars can easily cause rapid changes in wireless channel quality. Determining how to ensure extremely low or zero delays is also a very important issue.

\section{Conclusions}
We use network calculus to study the latency guarantee for UbiI in 6G. Specifically, we first propose the demands and challenges of 6G, EC, and edge DL, especially UbiI integrated by them. Then, we propose the system architecture and network model of UbiI and use network calculus to derive the upper bound of the E2E service delay of UbiI. In addition, we demonstrate the modeling process of the latency guarantee for UbiI with two case studies and verify the effectiveness of the network calculus approach to solve the latency guarantee for UbiI in 6G. Finally, some open issues of latency guarantee for UbiI in 6G are summarized for future work.

\section*{Acknowledgments}
This work was supported in part by the Natural Science Foundation of China (Nos. 61572191, 61602171 and 62072175) and the Natural Science Foundation of Hunan Province, China (No. 2020JJ4058).

\end{document}